\documentclass[amsmath,aps,prl,twocolumn,groupedaddress, nofootinbib, amssymb,floatfix,superscriptaddress]{revtex4-2}

\usepackage{physics,amsmath}
\usepackage{tabularx}
\usepackage{euscript}
\usepackage{epsfig,psfrag,subfigure}
\usepackage{dcolumn}
\usepackage{bm}
\usepackage{setspace}
\usepackage{color,soul,xcolor}
\usepackage{amsfonts}
\usepackage{exscale}
\usepackage{multirow}
\usepackage{hyperref}
\usepackage{wrapfig}
\hypersetup{colorlinks=false,linkcolor=black}
\usepackage[left]{lineno}
\usepackage[normalem]{ulem} 
\usepackage{placeins}

\begin{document}

\title{femto-PIXAR: a self-supervised neural network method for reconstructing femtosecond X-ray free electron laser pulses}

\author{Gesa Goetzke$^*$}
\affiliation{Deutsches Elektronen-Synchrotron (DESY), Notkestrasse 85, 22607 Hamburg, Germany}
\author{Rajan Plumley$^*$}
\affiliation{Department of Physics, Carnegie Mellon University, Pittsburgh, PA, USA}
\affiliation{Stanford Institute for Materials and Energy Sciences, Stanford University, Menlo Park, CA, USA}
\affiliation{SLAC National Accelerator Laboratory, Menlo Park, CA, USA}
\author{Gregor Hartmann}
\affiliation{Helmholtz-Zentrum Berlin für Materialien und Energie GmbH, Albert-Einstein-Strasse 15, 12489
Berlin, Germany}
\author{Tim Maxwell}
\affiliation{SLAC National Accelerator Laboratory, Menlo Park, CA, USA}
\author{Franz-Josef Decker}
\affiliation{SLAC National Accelerator Laboratory, Menlo Park, CA, USA}
\author{Alberto Lutman}
\affiliation{SLAC National Accelerator Laboratory, Menlo Park, CA, USA}
\author{Mike Dunne}
\affiliation{SLAC National Accelerator Laboratory, Menlo Park, CA, USA}
\author{Daniel Ratner$^{\dagger}$}
\affiliation{SLAC National Accelerator Laboratory, Menlo Park, CA, USA}
\author{Joshua J. Turner$^{\dagger}$}
\affiliation{Stanford Institute for Materials and Energy Sciences, Stanford University, Menlo Park, CA, USA}
\affiliation{SLAC National Accelerator Laboratory, Menlo Park, CA, USA}

\begin{abstract}
    X-ray Free Electron Lasers (X\nobreakdash-FELs) produce ultrafast pulses in a wide range of lasing configurations, supporting a broad variety of scientific applications including structural biology, materials science, and atomic and molecular physics.  Shot-by-shot characterization of the X-FEL pulses is crucial for analysis of experiments as well as for tuning the X-FEL performance. However, for the weak pulses found in advanced configurations, e.g. those needed for monochromatic, two-pulse studies of quantum materials, there is no current method for reliably resolving pulse profiles.
    Here we show that an interpretable neural network (NN) model can reconstruct the individual pulse power profiles for sub-picosecond pulse separation without the need for simulations. Using experimental data from weak X-FEL pulse pairs, we demonstrate a NN can learn the pulse characteristics on a shot-by-shot basis when conventional methods fail. This new method enables the characterization of weak pulses—a condition expected to dominate future pulse scenarios such as standard LCLS-II delivery—and opens the door to a wide range of new experiments.
\end{abstract}

\maketitle

\newpage
\def\thefootnote{*}\footnotetext{These authors contributed equally to this work}
\def\thefootnote{\textdagger}\footnotetext{E-mail: dratner@slac.stanford.edu, joshuat@slac.stanford.edu}

\section{Introduction}
Fluctuations in microscopic systems are directly related to their fundamental excitations -- a concept dating back to Einstein’s description of the Brownian motion of pollen particles. This relationship has motivated the need for measuring fluctuations within microscopic systems to directly compare with first-principles theoretical models. However, the spatial and temporal scales of quantum systems far exceed those accessible by classical microscopy, and despite the advances in modern light sources, the energy scales pertinent to these fluctuations remain challenging to probe. 
One notable exception to this has been in inelastic neutron scattering, where small energy changes of scattered neutrons can be observed, and has led to understanding of dispersion in a myriad of areas, from frustrated magnets and high temperature superconductors, to biological tissue and exotic topological structures, such as skyrmions. However, the reliance of inelastic neutron scattering on large crystals and its limitations with strongly absorbing elements restrict its applicability. 
The advent of the free electron laser has changed all of this. Its capability to generate femtosecond~(fs) laser pulses with extremely high pulse energies at X-ray wavelengths has ushered in a new era of science, with new developments rapidly evolving in many fields~\cite{Bostedt-2016-RMP, rossbach201910} including structural biology~\cite{Boutet-2012-Science, Chapman-2011-Nature}, enzyme catalysis~\cite{rose2021unprecedented}, and astrophysics~\cite{Bernitt-2012-Nature, Vinko-2012-Nature}. Furthermore, the development of monochromatic, two-pulse configurations brings new potential for fluctuation studies within quantum materials at their relevant timescales, i.e. at the meV energy scale and below~\cite{sun2020realizing, gutt2008measuring, plumley2024ultrafast,seaberg2017nanosecond, seaberg2021spontaneous, roseker2018towards}.

Importantly, with X-FELs it is now possible to deliver X-ray pulses at \textit{sub-picosecond} spacing for experiments, utilizing the transverse coherence of each X-FEL pulse to resolve instantaneous snapshots in structures with sub-Angstrom resolution. These stroboscopic scattering events are repeated and compared to measure fluctuations within the sample, offering unprecedented access to dynamic processes at ultrafast timescales. 
This highlights the urgent need for reliable characterization of X-FEL laser pulse intensities with femtosecond resolution.
Notably, X-FELs are highly sensitive to operational parameters which govern the durations, energies, and sequences of X-ray pulses~\cite{decker2022tunable, decker2022two}. Compounding this complexity, X-FELs exhibit intrinsic intensity fluctuations, causing stochastic variations in the initial pulse energy spectrum and photon density from shot to shot~\cite{bonifacio1994spectrum, sun2018pulse}. These beam intensity fluctuations from the X-FEL present significant challenges for experiments where accurate characterization of pulse properties, including pulse duration and energy density, are needed for shot-to-shot corrections in order to extract meaningful results from the measurement series.
Existing non-invasive diagnostics for absolute pulse energies like the gas energy monitor \cite{hau2010near} or the  X-ray gas monitor \cite{Tiedtke2008} cannot resolve the individual pulses at this few fs timescale. Other diagnostics, such as angular streaking \cite{Hartmann_2018, Li2018} can predict X-ray pulse profiles in the attosecond to few fs regime while THz streaking \cite{Hoffmann2018} has shown similar capabilities 

\begin{figure*}[tbh]
  \centering
  \includegraphics[width=\textwidth]{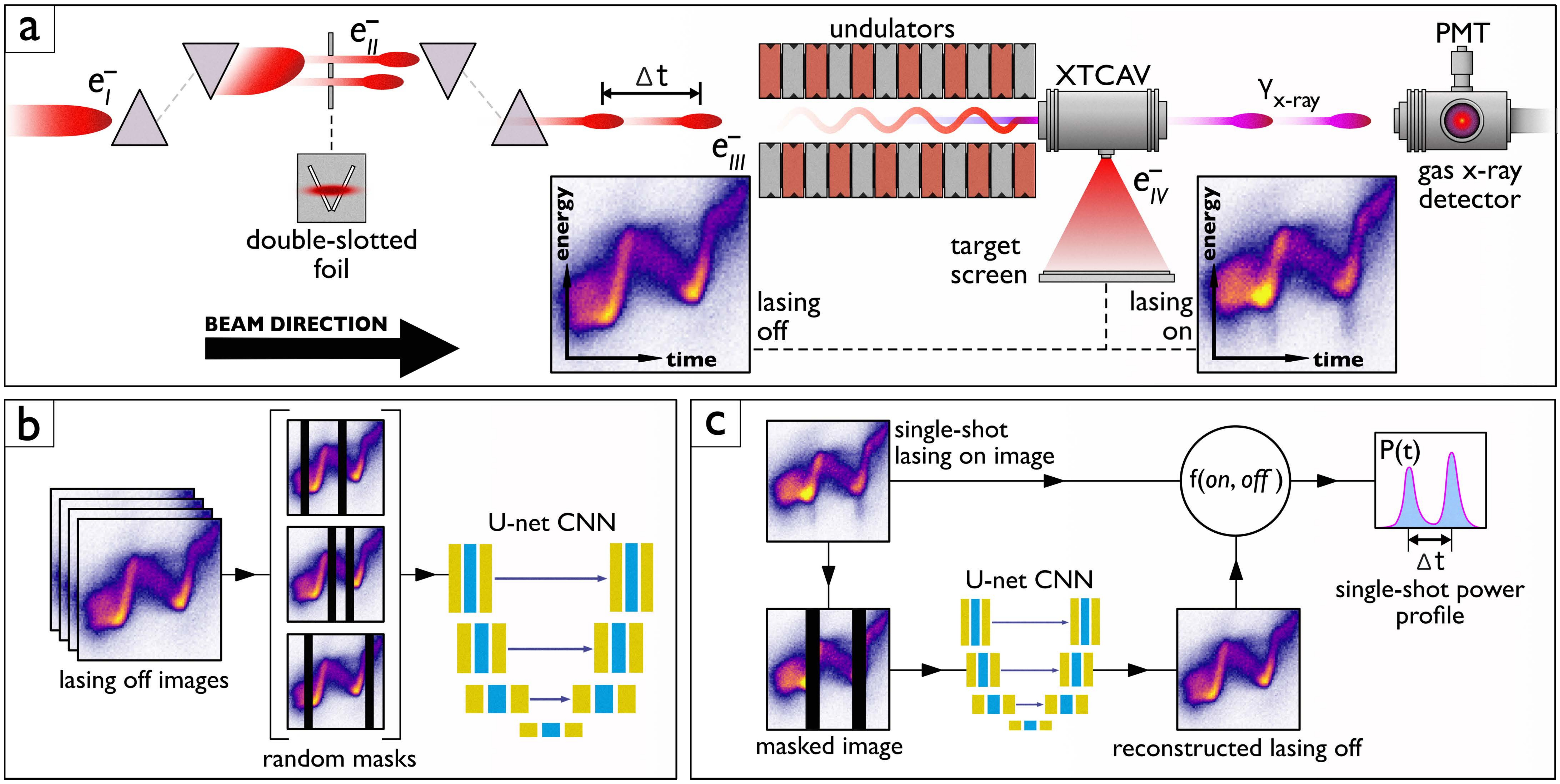}
  \caption{(a) Diagram of the generation and shaping of a single-shot X-ray pulse pair at the LCLS, beginning with the transmission of a single electron bunch through the double-slotted foil. Both, the slotted foil method for pulse separation and the XTCAV diagnostic are described in the text. (b) Diagram of the ML training process. (c) Diagram of the ML-based reconstruction of the transient X-ray power profile $P(t)$ from a single-shot lasing-on image.}
  \label{fig:scematics}
\end{figure*}

in the fs regime. 
However, these methods require sophisticated setups and can be considered complex experiments themselves rather than a standard diagnostic at this time.
Sanchez-Gonzalez et al. \cite{sanchez2017accurate} utilized machine learning to combine information from different detectors in order to improve pulse diagnostics, including the temporal distance of two electron bunches, but did not provide pulse intensity information on these timescales.
A method that was successfully used to analyze pulses in the few fs timescale is the XTCAV method \cite{Ding2015, MarTwinBunch, decker2022tunable}, described in more detail in our setup, and data and methodology chapter.  The XTCAV pulse monitoring system works reliably for typical self-amplified spontaneous emission (SASE) experiments and has become a standard operational pulse diagnostic for X-FELs such as the LCLS.
However, advanced X-FEL modes, e.g. self-seeding to provide a monochromatic beam, results in weak pulses that the conventional XTCAV analysis cannot resolve. Moreover, these modes are expected to become the standard for next generation high-repetition-rate XFELs, including LCLS-II. For these modes, the existing XTCAV analysis fails to resolve the signal over the background.
These challenges have so far hindered use of this new capability to deliver new science in the observation of fluctuations for understanding properties in materials.

We address this need by developing a novel self-supervised machine learning (ML) framework called femto-PIXAR: \textit{femtosecond Power Inference of X-ray pulses using AI References}, which provides sufficient sensitivity to resolve microjoule-scale pulses with femtosecond pulse separation. We show our approach accurately predicts pulse energies for an experimental dataset using a self-seeded beam, verified against total pulse energy monitors in real X-FEL data, and confirms that the predictions match the expected pulse delays. This demonstration will make ultrafast fluctuation studies possible in the sub-picosecond regime, and will also provide a tool for a wide array of X-FEL experiments in many other fields as well. 
\FloatBarrier
\section{Setup}
For this experiment, the X-FEL was set up to produce two monochromatic X-ray pulses starting from a single, long electron bunch, itself generated from two overlapping injector lasers. The two X-ray pulses were created using a double-slotted V-shaped foil (for time domain control) \cite{Emma2004} \cite{Ding2015}, and hard X-ray self-seeding \cite{Geloni_2011, Amann2012, Lutman_2014} (for longitudinal coherence of each individual pulse and energy selection). To our knowledge this configuration, depicted in Fig.\,\ref{fig:scematics} a), has never been used before. The experiment was conducted at a photon energy of 8.35\,keV (Ni $K$-edge), with slotted-foil pulse separations of $0\,\mathrm{fs}$ (single pulse), $\approx 20\,\mathrm{fs}$, and $\approx 30\,\mathrm{fs}$. Different stages of the electron bunch shaping sequence are labeled $e^{-}_{\text{I}}$, $e^{-}_{\text{II}}$, $e^{-}_{\text{III}}$, $e^{-}_{\text{IV}}$, representing the same electron distribution at different instances as it proceeds in time. A 3\,$\mu$m thick aluminum foil with two slots of variable separation from 1\,mm to 5\,mm was placed in the middle of the dispersive section of the bunch compressor BC2 \cite{Emma2004}. This is because a bunch compressor in a FEL works similarly to a chirped pulse amplification laser system: an initially chirped pulse is compressed due to different path lengths for different energies, producing a spatial expansion of the electron beam at the position of the slotted foil.
At the locations where the electron bunch $e^{-}_{\text{I}}$ passes through the foil the electron emittance was `spoiled' (i.e. increased) due to coulomb scattering. The emittance growth suppresses FEL amplification in the undulator, so that only the electrons $e^{-}_{\text{II}}$ that pass through the slits of the V-shaped foil will later produce X-ray pulses. 
Downstream of the bunch compressor the relative distance between the unspoiled electron distributions $e^{-}_{\text{III}}$ is in the direction of travel, resulting in an arrival time difference $\Delta t$,  that can be changed by shifting the position of the slotted foil with respect to the beam center line. $e^{-}_{\text{IV}}$ passes through the undulator magnets resulting in X-ray lasing. Between the first and second undulator segment a hard X-ray self-seeding scheme is applied in order to obtain monochromatic photon pulses (not shown here). Downstream of the undulators, in the X-band Transverse-deflecting-mode CAVity (XTCAV) \cite{Ding2011, Behrens2014}, a radio-frequency electromagnetic field streaks the electron beam. The XTCAV is combined with an energy spectrometer that provides energy dispersion perpendicular to the streaking direction. The electron bunch is deflected onto a fluorescence target screen, dispersing the electron energy across the vertical axis while the arrival time is dispersed along the horizontal axis. The fluorescence is recorded with a CCD camera, allowing measurement of the electron bunch's longitudinal energy profile with femtosecond resolution. Examples of XTCAV images are displayed in Fig.\ref{fig:scematics}. This transient electron energy profile can be analyzed with or without X-ray lasing suppressed in order to determine the power profile of the outgoing X-ray laser as a function of time.  This highly sensitive comparison  of lasing-on and lasing-off beams is at the center of the analysis presented in this article.
The X-ray laser pulses emitted by the electron bunch are unaffected by the XTCAV deflector magnet and continue downstream to the X-ray gas energy monitor (GEM), where the total X-ray pulse energy is transmissively monitored on a shot-to-shot basis using a photomultiplier tube (PMT) to measure the fluorescence induced in a small volume of N$_2$ gas~\cite{hau2010near}.
After the GEM, the X-rays continue down the beamline to the experimental endstations. This entire process repeats at a 120~Hz repetition rate.
\begin{figure*}[bth]
\centering
\includegraphics[width=0.95\textwidth]{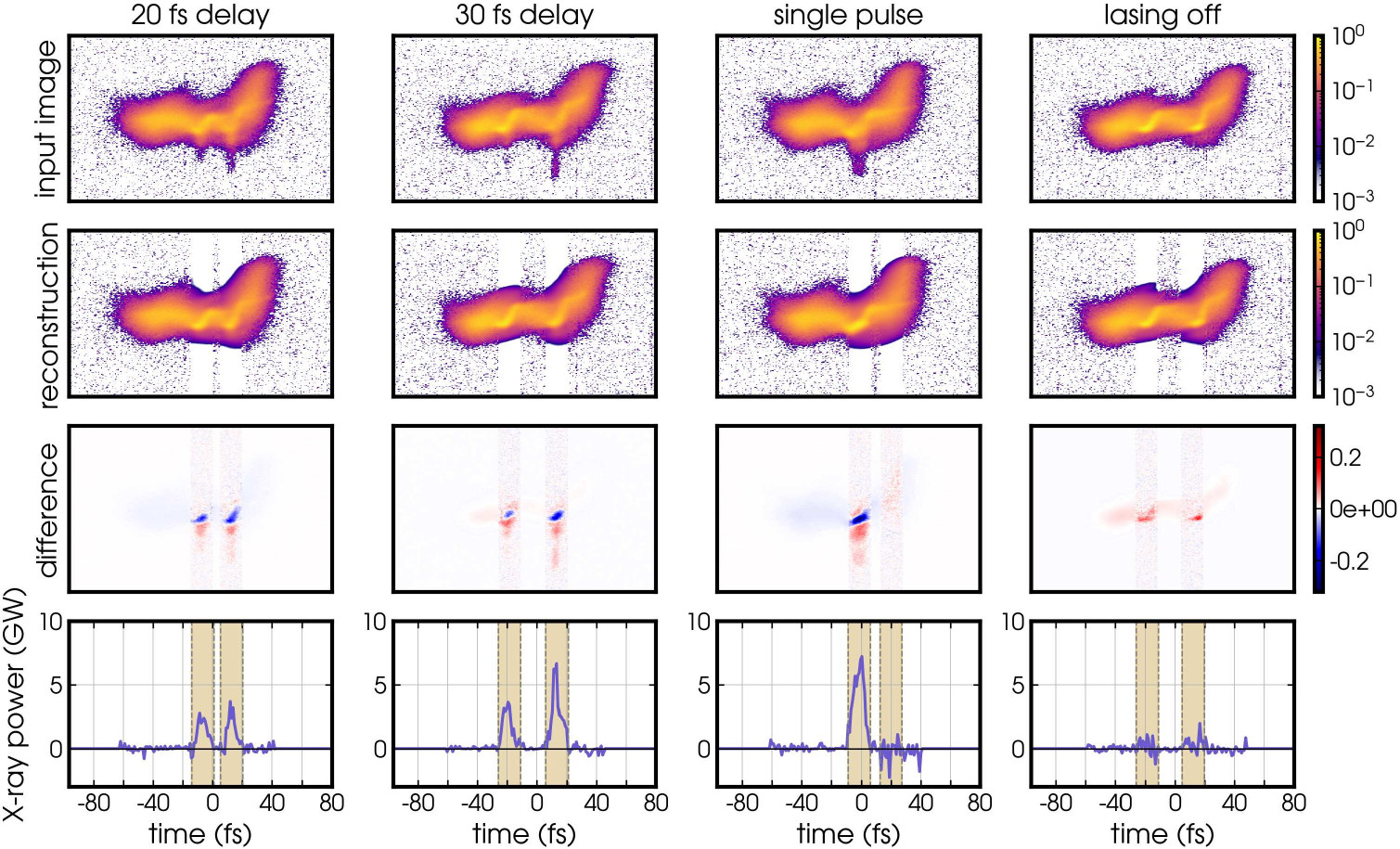}
\caption{Demonstration of femto-PIXAR reference creation: The XTCAV images show the respective longitudinal electron densities of the FEL bunch, with the x-axis corresponding to time and the y-axis corresponding to the electrons’ kinetic energy. The color density in each panel represents the amount of electron charge per pixel. The top row displays background-subtracted XTCAV images for three different slotted-foil setups (lasing-on) plus a lasing-off sample at a 30 fs delay. The second row shows the predicted lasing-off reconstruction; the third row the difference between the measured XTCAV data and that reconstruction; and the bottom row the resulting reconstructed power profiles. }
\label{fig:demoSamples}
\end{figure*}

\section{Data and Methodology}
As described in the previous chapter, the XTCAV system images the longitudinal (time-energy) distribution of the electron bunch with or without lasing suppressed. By comparison with a lasing-off reference, it is possible to infer the temporal profile of the X-ray pulse. However, because each shot is unique and the lasing-on and lasing-off references cannot be measured simultaneously, this poses a fundamental challenge in determining the change in the electron phase space due to the lasing process. The ``classical'' approach has been to record a set of images while suppressing lasing, cluster these images, and then select the most similar cluster to reference against each lasing-on image \cite{Ding2011, Maxwell2014, Behrens2014}. While the classical approach has been demonstrated for various two-pulse modes \cite{Ding2015, MarTwinBunch, decker2022tunable}, it struggles for small pulse energies, when shot-to-shot differences between the true lasing-off electron bunch and the selected reference cluster exceed the lasing-induced changes.  As a result, when combining self-seeding with two-pulse modes, the classical method breaks down.

A deep learning method proposed by Ren et al.~\cite{Ren2020} sidesteps the need for matched lasing-off references by training a neural network (NN) directly to predict power from a lasing-on image; however, the NN in ~\cite{Ren2020} relies on simulated labels for training, and it is particularly challenging to simulate the complex electron beam distribution produced by the slotted foil. 
For the slotted foil setup used in this work, only a narrow segment of the beam lases.  We can exploit this feature to approximate a reference in the short lasing region by using information in the nearby non-lasing regions.  For example, Zeng et al. \cite{zeng2022online} suggested to use a polynomial regression across the lasing region as a reference. However, the polynomials cannot model the complex phase space of this setup and fail to reproduce the weak lasing pulses.

Here, we instead propose using a NN to generate the full corresponding lasing-off reference image. We employ a U-net architecture \cite{Ronneberger2015}, originally developed for biomedical image segmentation, and now broadly used in different fields \cite{reviewU, Lee2023, Gorse2024, KAMALI2024}. A U-net uses an encoder–decoder architecture with symmetric skip connections, letting it capture the global structure within the data while preserving fine details, making it ideal for resolving low signal-to-background features. We trained the NN using self-supervision, eliminating the need for simulations. Specifically, we masked regions of lasing-off pulses, and trained the NN to regenerate those regions. Once trained, the network receives a lasing-on image with masked lasing regions and reconstructs the best lasing-off reference. We then calculated the center of mass (COM) to retrieve the X-ray pulse distribution \cite{Behrens2014}.
Our method provides high-fidelity lasing-off references without relying on the accuracy of computationally-expensive simulations while retaining interpretability through direct observation of the generated reference.

Fig.\,\ref{fig:scematics} b) and c) shows the core steps for the network training and inference process.  A technical description of the steps is included in the supplemental material. The network was trained on about 30k measured lasing-off samples with an additional 1.5k frames being used as validation to check for overfitting. Additionally, we have reserved 460 lasing off samples as a test dataset, that were never seen by the network.
The inputs to the network were cropped, background subtracted and normalized lasing off images, where all values within the two random mask-regions were set to zero (black bars in Fig.\,\ref{fig:scematics} b) and c)). The width of an individual mask was chosen to be 16 pixels and the distance range for the two masks was set from 0 to 40 pixels, corresponding to approx.  0 to 38~fs. We then trained a U-net to reconstruct the electron phase space of the lasing-off data (Fig.\,\ref{fig:scematics} b) ). As the center of mass is the critical value for evaluation, we used a weighted combination of the mean squared error (MSE) of pixel values and the absolute error of the COM. The COM has a relative low weight of 1e-4 as we found that higher weights significantly worsen the reconstruction. After training, we masked the lasing regions in a lasing-on data set. As the network is trained only on the lasing-off data, it reconstructs the best matching lasing-off electron phase space exactly for a given image (Fig.\,\ref{fig:scematics} c). This lasing-off image will also exactly match the original $x$ and $y$ position, so no shifting in the time axis or normalization with the GEM (as in the classical approach) is necessary. 
A crucial step of this process was to determine the correct mask positions in the lasing data. The separation of the two pulses was approximately fixed by the slit separation, but due to shot-to-shot differences in electron bunch energy and compression of the electron bunch, the position of the features relative to the center of mass of the electron bunch was not constant.  Our method scans the image to find two locations where input and reconstruction differ most, centering the masks on the maximum signal. This procedure is described in more detail in the supplementary material. We then use the methods from Behrens et. al \cite{Behrens2014}, marked as $f(\text{on},\text{off})$ in Fig.\,\ref{fig:scematics} which calculates the transient X-ray power profile by using the COM from the lasing-on image ``on'' and subtracting the COM of the network reconstructed lasing-off image ``off''.
\FloatBarrier
\pagebreak
\section{Results}

\FloatBarrier

Fig.\,\ref{fig:demoSamples} shows example reconstructions for the three different configurations together with one lasing-off example.  All images were randomly drawn from the top 1\% of shots with the largest GEM values, ensuring a clearly visible signal; none were used for network training.
The reconstructions reliably reproduce the two-dimensional structure, and the main difference between lasing on and lasing off is within the masked regions. We note that the U-Net effectively de-noises the reconstructed masked region, as the random noise signals cannot be inferred from the neighboring regions. However, this has no impact on the pulse reconstruction. While most of the difference between lasing on and lasing off is within the masked region, there is some residual signal in the regions outside the masks, where the network has full information and should be able to reconstruct the beam perfectly.

\begin{figure*}[hb]
\centering
\includegraphics[width=\textwidth]{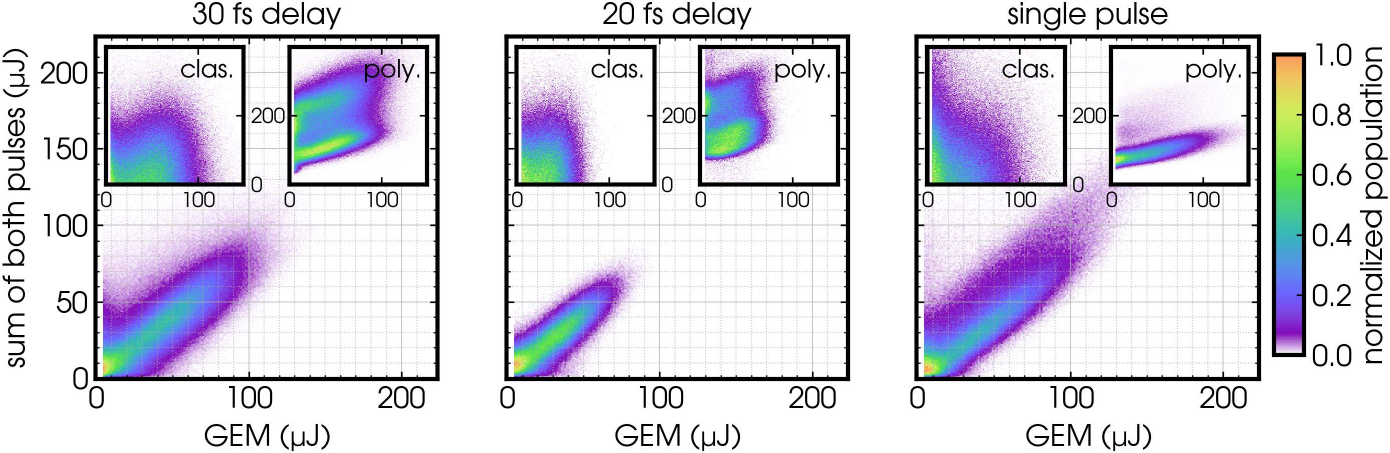}
\caption{Correlations with the GEM for the three different delays using the U-net approach. Both the classical method (clas.)~\cite{Ding2011} and polynomial regression (poly.)~\cite{zeng2022online} reconstruction methods displayed in the insets show poor correlation between the predicted power-profile integrals and the total GEM pulse energy especially on the weaker double pulses ( $\approx 20\,\mathrm{fs}$ and $\approx 30\,\mathrm{fs}$ delay).}
\label{fig:gmd}
\end{figure*}

 We find this residual increases when the center-of-mass (COM) loss is introduced or assigned greater weight. In principle the effect could be removed through additional hyper-parameter tuning; however, because these minor artifacts do not affect the extracted power profiles, no further optimizations were necessary. To demonstrate the efficacy of the proposed method, we benchmark the predicted power and delay against existing diagnostics. First, we compare the integral of the predicted profile with the total energy of both pulses measured by the GEM, which cannot resolve individual pulses on the fs scale but measures the total energy for each pair.  In Fig.\,\ref{fig:gmd} one can see the correlation of the integral of the reconstructed power profile with the GEM for the three different slotted foil configurations.
We note that while there is a strong linear correlation between the predicted power and GEM monitor, there is a disagreement in proportionality. Further experiments are necessary to determine if the disagreement is due to either the GEM or XTCAV calibration, or bias from the method. However, if the absolute pulse energy is a measurement of interest, it would be possible to correct for any proportional bias from the XTCAV diagnostic by using the GEM, though this change would not affect the experimental analysis outlined here.

Next, we compare our predictions to two existing state-of-the-art XTCAV analysis methods. In order to evaluate the correlation with the GEM, in the classical approach (see the previous section), the energy jitter (shift in the y-axis) cannot be compensated for by normalization with the GEM but must instead be corrected using a scheme that matches the head and tail of the COM profiles for the lasing-on and lasing-off images. We also compare our results with the polynomial regression method of Zeng et al. \cite{zeng2022online}. Some example predictions of this method can be found in the supplemental material. Fig.\,\ref{fig:gmd} compares the GEM correlation for all three methods across all three delays, and neither of the existing methods results in the expected correlation.
\onecolumngrid
\vspace{0.8cm}

\FloatBarrier
\clearpage

\begin{figure*}[hbt]
\centering
\includegraphics[width=0.9999\textwidth]{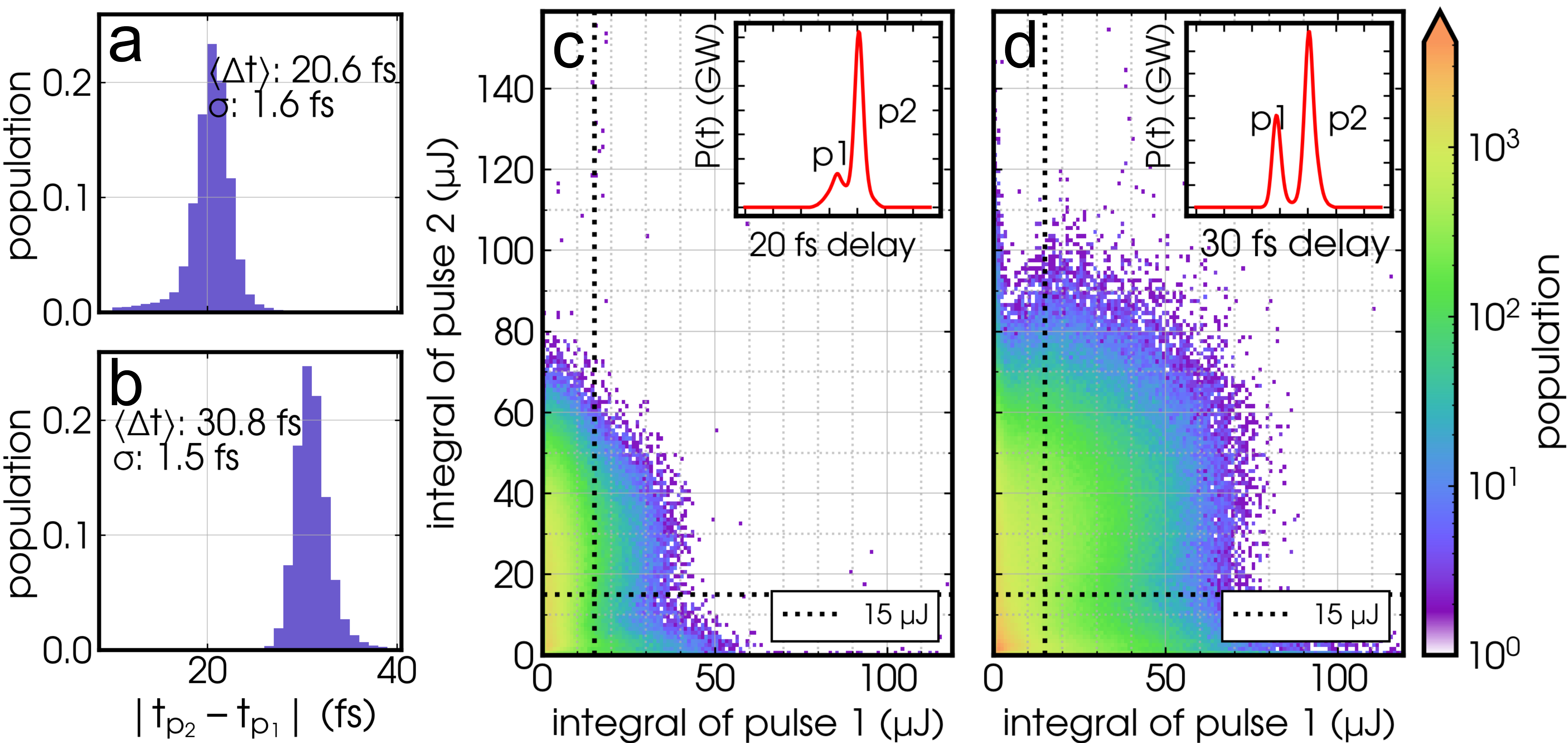}
\caption{(a) \& (b): Histograms of temporal separation $\Delta t = |t_{p_2} - t_{p_1}|$ between each pulse for the 20~fs and 30~fs delay setups, respectively. (c) \& (d): Histograms of the integrated energy in each pulse for both setups. The dashed line indicates the 15\,${\mu}$J minimum energy threshold used to select shots for the time-separation analysis in (a) \& (b). The insets show the average of the transient X-ray power profiles as as function of time $t$ (fs), where the relative average amplitudes of the pulse 1 (p1) and pulse 2 (p2) are indicated.}
\label{fig:amplitudeRatio}
\end{figure*}

\FloatBarrier 
\twocolumngrid
Fig.\,\ref{fig:amplitudeRatio} shows the temporal and energy characteristics of the X-ray laser pulses as retrieved by the U-net reconstruction scheme.  The time assignments of each shot were obtained by fitting the X-ray power trace with two Gaussian functions.  For these plots we took only the data where both Gaussian curve fits indicated a signal larger than 15\,${\mu}$J.  We find that the predicted time difference between both pulses closely matches the delay given by the slotted foil condition. The temporal separation jitter aligns well with the 9$\%$ jitter observed by Ding et al. \cite{Ding2015}.  As an additional benchmark we checked the predicted pulse energies for lasing-off data with randomly placed masks and find that 80$\%$ of the data has a deviation from zero of less than 15 $\mu$J, and 99.8$\%$ less than 25 $\mu$J. We find that the pulse energy difference between the two pulses is centered well around zero for lasing-off-data, indicating that our approach does not introduce a bias in the ratio between the first (p1) and the second (p2) pulse. We also compute the relative squared error between the input frames and the lasing off frames for this data where we have the ground truth (true signal). More information on these additional benchmarks showcasing the validity of the method can be found in the supplemental material.

\section{Discussions and Summary}
In conclusion, femto-PIXAR is a novel X-FEL diagnostic method which we demonstrate by reconstructing X-ray power profiles from real X-FEL data. For the configuration used to deliver monochromatic pulse pairs at few fs distance, the classical method of XTCAV analysis, which uses a matching algorithm to subtract similar lasing-off shots for the energy loss calculation, fails to reproduce meaningful correlations with the X-ray gas energy monitor. By contrast, our method is able to distinguish the relative amplitude of each X-ray pulse on a shot-to-shot basis which is necessary for X-ray photon fluctuation spectroscopy experiments~\cite{Shen-mrsa-2021} and other future experiments at X-FEL facilities that utilize non-standard beam setups. Our approach has various benefits, avoiding the need for energy calibration with an external reference. 
A neural network based scheme also scales well for the large datasets expected from high-throughput experiments enabled by next generation X-FEL facilities such as the LCLS-II~\cite{galayda2018lcls}. As opposed to previous deep-learning work, our approach uses self-supervision, avoiding the need for high-fidelity simulations. Our method also reconstructs the full 2D phase space, providing interpretability of the network and giving crucial information about when the network might not be applicable for a specific dataset.
 Future work will consider extensions to more general X-FEL configurations beyond the slotted foil. 
For example, using other diagnostics in addition to the XTCAV could enable analysis of configurations with lasing across the full electron bunch \cite{sanchez2017accurate}, e.g. to provide the femtosecond resolution needed for temporal ghost imaging \cite{ratner2019pump} 
or help overcome hardware limitations that restrict temporal resolution to a few femtoseconds,  enabling access to the attosecond regime.
The analysis chain after reconstructing the lasing-off reference can also make use of machine learning, e.g. incorporating spectral measurements to refine the X-ray pulse profile \cite{ratner2021recovering}.

\section{Funding}
U.S. Department of Energy (DE-AC02-76SF00515., DE-SC0022216); IVF project InternLabs (HIR3X). (0011193).

\section{Acknowledgment}
Use of the Linac Coherent Light Source (LCLS), SLAC National Accelerator Laboratory, is supported by the US Department of Energy, Office of Science, Office of Basic Energy Sciences under Contract No. DE-AC02-76SF00515.  This work has in part been funded by the IVF project InternLabs-0011 (HIR3X). R. Plumley acknowledges support from the US Department of Energy, Office of Science, Basic Energy Sciences, for the Materials Sciences and Engineering Division under contract DE-AC02-76SF00515. J.J.Turner acknowledges support from the U.S. Department of Energy (DOE), Office of Science, Basic Energy Sciences under Award No. DE-SC0022216. D. Ratner acknowledges support from the U.S. Department of Energy (DOE), Office of Science, Basic Energy Sciences under Award No. DE-AC02- 76SF00515.

\section{Disclosures}
The authors declare no conflicts of interest.

\section{Data Availability}
The data presented in this study are available on request from the corresponding authors.

\section{Supplemental Materials}
See Supplement 1 for supporting content.

\vspace{1cm}

\bibliography{lib.bib}

\pagebreak
\setcounter{equation}{0}
\setcounter{figure}{0}
\setcounter{table}{0}
\setcounter{page}{1}
\makeatletter
\renewcommand{\theequation}{S\arabic{equation}}
\renewcommand{\thefigure}{A\arabic{figure}}

\onecolumngrid

\begin{center}
\textbf{\large Supplemental Materials}
\end{center}

\section{A1: Data preparation}
The centered XTCAV images have a shape of 240 x 240 pixel. They are centered in both dimensions using the center of mass. A median dark image is subtracted for all the analysis conducted in this paper. Before the data is fed into the network, it is normalized to have intensities roughly between 0 and 1. The reconstruction of the network is re-normalized again to the original range of values to ensure an analogous data handling to the other approaches.
Before obtaining the power profiles, a median filter is used to identify the signal region and set the other regions to zero. This approach is just used to identify the signal region, no median operation is applied to the image. 

\section{A2: Obtaining the X-ray power profile from the XTCAV spectrograph}
The time-dependent power profile $P(t)$ of the X-ray pulse can be derived by the following expressions from FEL theory~\cite{Ding2011, Behrens2014}.
\begin{equation}
    P(t) = \frac{I(t)}{e}(\langle E_{\text{off}}(t)\rangle - \langle E_{\text{on}}(t)\rangle),
\end{equation}
and
\begin{equation}
    P(t) \propto  I(t)^{2/3}(\sigma^{2}_{\text{on}}(t) - \sigma^{2}_{\text{off}}(t)),
\end{equation}
where $I(t)$ is the beam current, $e^{-}$ the electron charge,  $\langle E_{\text{off}}(t)\rangle$ and $\sigma_{\text{off}}(t)$ are the time-dependent $e^{-}$ energy expectation value and variance for the case when the energy distribution of the electrons in the electron bunch was not influenced by lasing, and $\langle E_{\text{on}}(t)\rangle$ and $\sigma_{\text{on}}(t)$ for the case when the electron distribution was shaped by the lasing process.  (S2) can only give proportionality and requires power normalization with the GEM. As this is the value we want to benchmark with, we decided to use equation (S1) in order to obtain the power profiles.

\section{A3: Network analysis}
We used an U-net architecture as described in \cite{Ronneberger2015}. The optimizer is RMSprop of \href{https://github.com/pytorch/pytorch}{pytorch} with a weight decay of 1e-8. Based on a few manual trials, we chose a learning rate of $1\times10^{-4}$ and a batch size of~32.  A more extensive hyper-parameter search might yield slightly better reconstructions or faster convergence. 

For the loss function~$L$, we used a weighted sum comprising the mean-squared error between the reconstruction image~$r$ and the input image~$i$, both containing $240^{2}$ pixels, combined with the mean absolute error of the centers of mass (COM), denoted by $com_i$ and $com_r$:
\begin{equation}
L = \frac{1}{240^{2}}\sum_j (i_j - r_j)^{2} + w_{COM}\,\langle |com_i - com_r| \rangle
\end{equation}
The weight of the COM term, $w_{COM}$, was set to the relatively low value of $1\times10^{-4}$, as higher weights significantly degraded the reconstruction.\\

X-FEL shots where essentially no lasing occurs with $E_{\text{GEM}}< 5\,\mu$J are included in the lasing-off dataset to get more training data. This results in about 30k lasing-off samples that are used for training, with an additional 1.5k frames being used as validation dataset to observe for overfitting. Additionally, 460 lasing-off samples were held out as a test set; they were not exposed to the network during training and were used only for the residual-error benchmark.  
In total, our lasing-on dataset consists of about $4\times10^5$ samples with approximately $20$\,fs difference, about $1.5\times10^6$ samples with approximately $30$\,fs difference and $9\times10^4$ with only one pulse.
Training was stopped when the validation loss failed to decrease by at least $7.5\times10^{-7}$ within a patience window of 25 steps.  
On an RTX-2080 S GPU this required roughly 3 hours without any specific optimization.  
We do not expect a network trained on one dataset to generalize well enough to handle a dataset originating from a completely new FEL setup. However, training is sufficiently fast that retraining a network for a new setup should be straightforward.

\section{A4: Masking procedure}
\begin{figure*}[h!bt]
\centering
\includegraphics[width=0.99\textwidth]{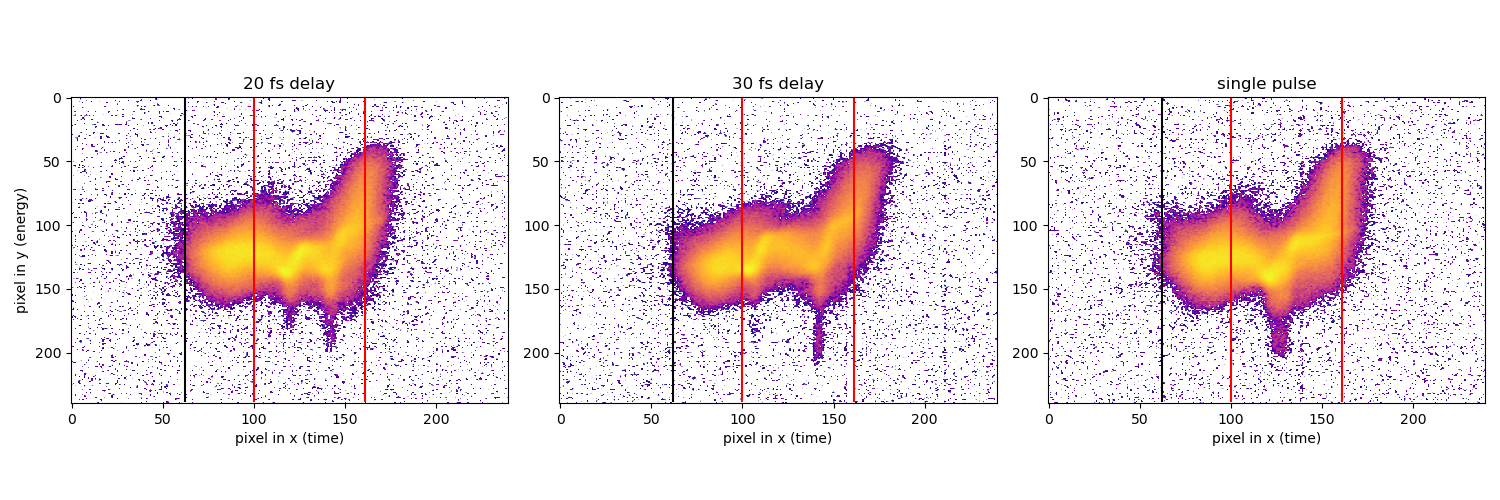}
\caption{The red lines mark the borders of the possible masking region during inference, and the black line marks the left border of the dummy mask, which is intentionally placed outside the lasing region during the search for the correct mask positions.}
\label{fig:maskingPosition}
\end{figure*}

During training, masks are placed at random positions to teach the network to reconstruct missing regions at various positions.  Because we wish to mask only regions where the electron-bunch signal is present, we restrict the masking area (to pixel 60–180, see Fig. \ref{fig:maskingPosition}).  
We further constrain the random masks so that the minimum distance between them is~0 (i.e.\ they do not overlap) and the maximum distance is 40 pixels.  
The mask size is fixed at 16 pixels; however, the exact size is not critical for the reconstruction.  
A discussion of the influence of mask size on the predicted pulse profiles and pulse separation follows below. 

During inference (i.e. when the trained network is used for analysis), the masks must be positioned so that they cover the lasing regions.  We let the network determine these positions directly. To accelerate the mask search, we do not search the entire 240-pixel frame but instead define a range where the lasing feature is expected.  
 We explicitly exclude the borders of the electron bunch, because shot-to-shot variations are greatest there and cannot be reliably inferred from the unmasked part.  
This masking area extends from pixel~100 to pixel~161 (see Fig.~\ref{fig:maskingPosition}). 
We then place one mask at a dummy position outside the expected lasing region (pixel~62).  This was a position that was also used during network training.
This step is necessary because the network was always trained with two masks; as expected, artifacts appear in the reconstruction if only a single mask is used.

We then construct a set of images, each containing the dummy mask plus a candidate position for the second mask, with every possible candidate position included in the set. The NN then processes the entire set. 
We identify the input mask position that produces the largest energy difference (mask~A) and the position with the second-largest difference that is at least one mask width away (mask~B), ensuring the masks do not overlap.  For further evaluation, the mask positions are ordered so that \emph{mask1} is the leftmost of mask~A and mask~B. These two masks are then fed into the network to obtain the final reconstruction.

\subsection{A5: Influence of the mask size on GEM correlation and time distribution}
\begin{figure*}[h!bt]
\centering
\includegraphics[width=0.99\textwidth]{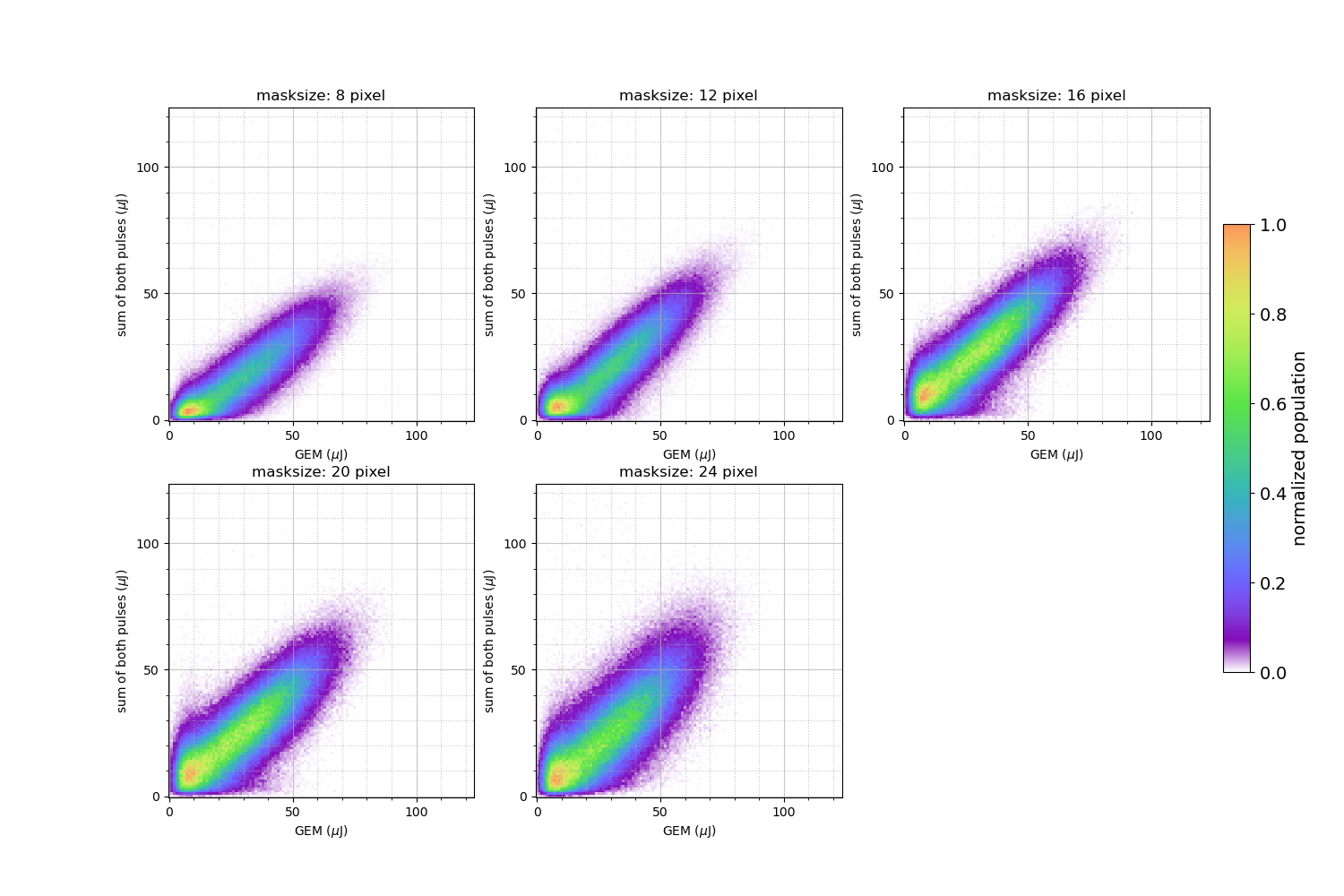}
\caption{Plots of GEM correlations for different mask sizes. The correlation is stable around a mask size of 16 pixels.  Each 2D histogram is normalized independently so that its maximum value is~1.}
\label{fig:gmdMasksize}
\end{figure*}

\begin{figure*}[h!bt]
\centering
\includegraphics[width=0.89\textwidth]{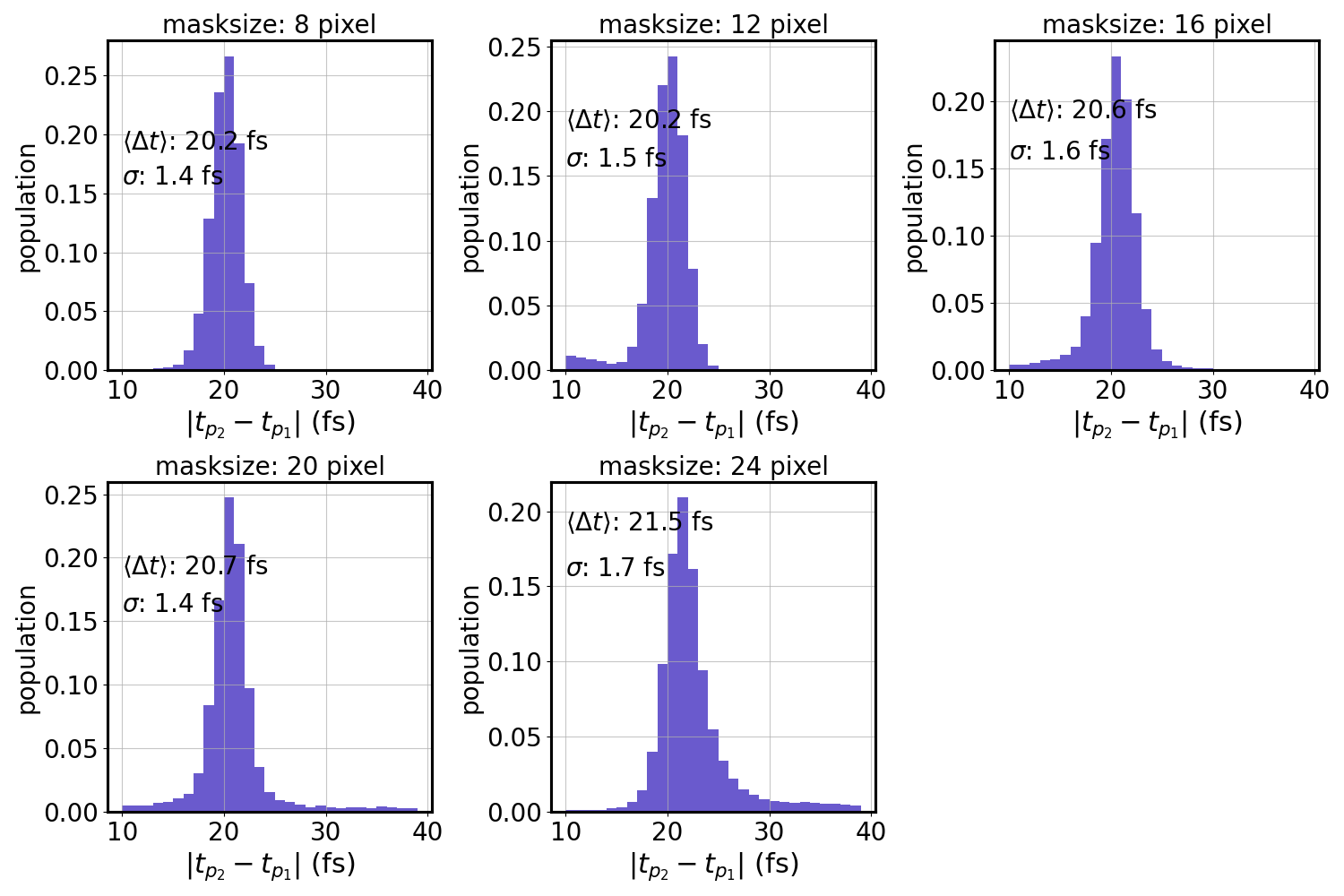}
\caption{Distribution of the difference in pulse arrival times for both two-pulse setups and various mask sizes.}
\label{fig:delTMasksize}
\end{figure*}

We chose a mask size of 16 pixels based on the extent required to cover the entire lasing feature of a strong photon pulse.  
To assess the influence of mask size, we trained and evaluated four additional networks identical in all respects except for the mask size, which was set to 8, 12, 20, and 24 pixels, respectively.  
The results on the 20 fs separation dataset are shown in Fig.~\ref{fig:gmdMasksize} and Fig.~\ref{fig:delTMasksize}.

Fig.~\ref{fig:gmdMasksize} shows that with a mask size of 8 pixels, the predicted photon pulse energies are lower because not all signal is captured. For a mask size of 24 pixels, the distribution is noticeably broader due to the inclusion of a wider noise background around the lasing region.  However, near 16 pixels the exact mask size appears quite robust and does not substantially affect the correlation.

Similarly, the predicted difference in arrival time is not strongly influenced by mask size. Only for the largest mask size (24 pixels) is a significantly higher arrival-time difference predicted.  

A larger mask size increases the risk that inaccuracies in the lasing-off electron-bunch energy distribution outweighs the actual lasing signal, leading to an inaccurate pulse prediction.

In summary, it is very straight-forward to find a matching mask size by inspection of the strongest pulses and does not require precise tuning for accurate reconstructions.  We found that our networks results are robust against different mask sizes as long as they are in a reasonable range.

\FloatBarrier
\section{A6: Classical analysis}
For the classical analysis the signal region is identified as described in the chapter data preparation, and then divided into 120 slices so that all lasing-on and lasing-off data has the same length in slices and can be compared directly. This method of finding the signal region and slicing also solves the challenge of aligning lasing on and lasing off horizontally (in the time domain), something that is necessary due to timing jitter.
 \href{https://scikit-learn.org/0.15/modules/generated/sklearn.cluster.AgglomerativeClustering.html}{HierarchicalClustering} with euclidean distance is used to cluster the projection on the y-axis (the time axis), into 500 groups. The projection on the y-axis corresponds to the electronic current if the signal is normalized with the total charge in the electron bunch, measured by an independent detector at the beam dump.
Normally, a few hundred lasing-offs are used to build a reference set, but to have a fairer comparison all lasing-off images that are used in the training of the network are provided to the algorithm. The COM profiles of each of these groups are averaged and used as reference. \\
In order to find the best matching reference group, the $y$-projections of a respective shot is compared to each of the $y$-projections of the lasing-off clusters.  This is done with the Pearson product-moment correlation coefficients. The non-diagonal elements of this matrix form an array where each element is the correlation coefficient between the lasing on current profile and each lasing off current profiles. This gives a measure how strongly the fixed lasing on data point  and each  lasing off reference current profile are related. The cluster with the strongest correlation is taken as the lasing off reference for this specific data point.

\section{A7: Implementation of polynomial regression method}
\begin{figure*}[h]
    \centering
    \includegraphics[width=0.8999\textwidth]{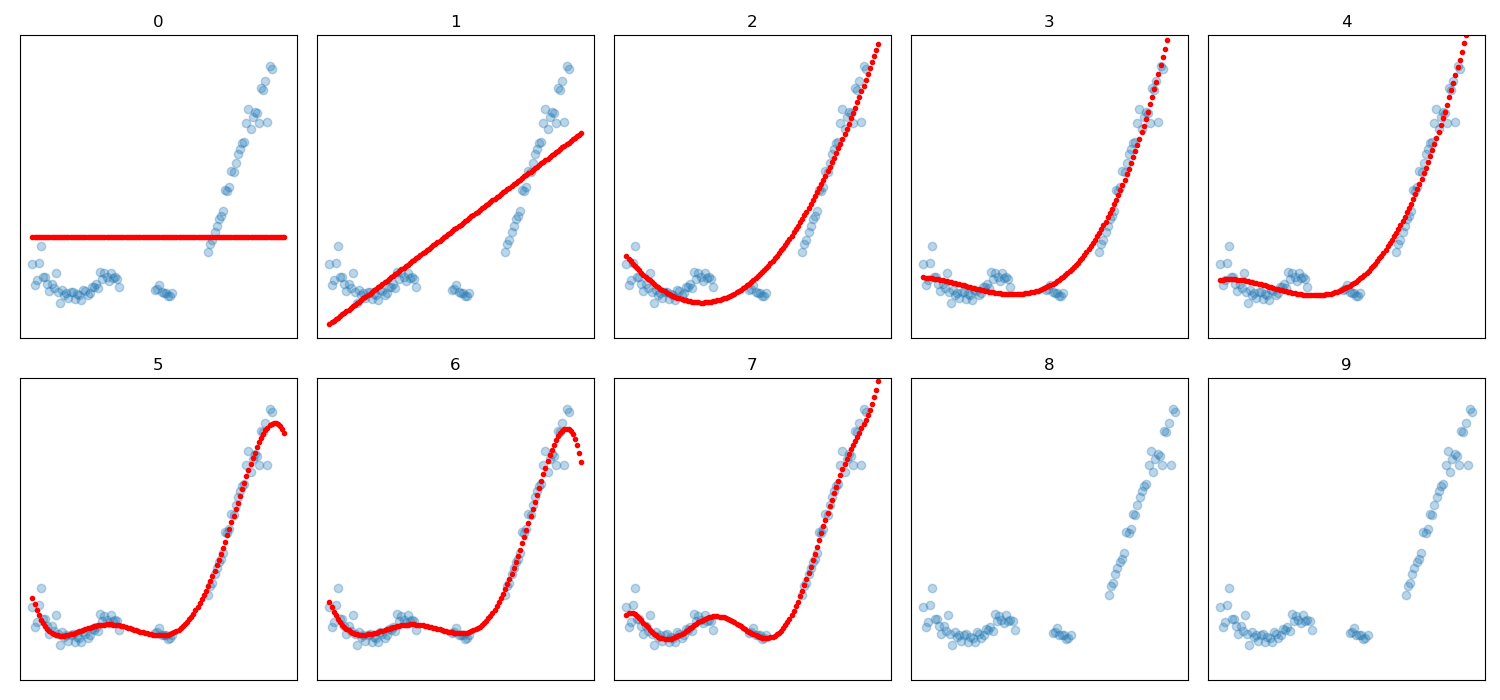}
    \caption{The blue dots mark the COM for one sample. The COM within the mask region is reconstructed with increasing polynomial degrees. For a degree higher larger seven, it does not converge.}
    \label{fig:polyDegree}
\end{figure*}

\begin{figure*}[h]
    \centering
    \includegraphics[width=0.8999\textwidth]{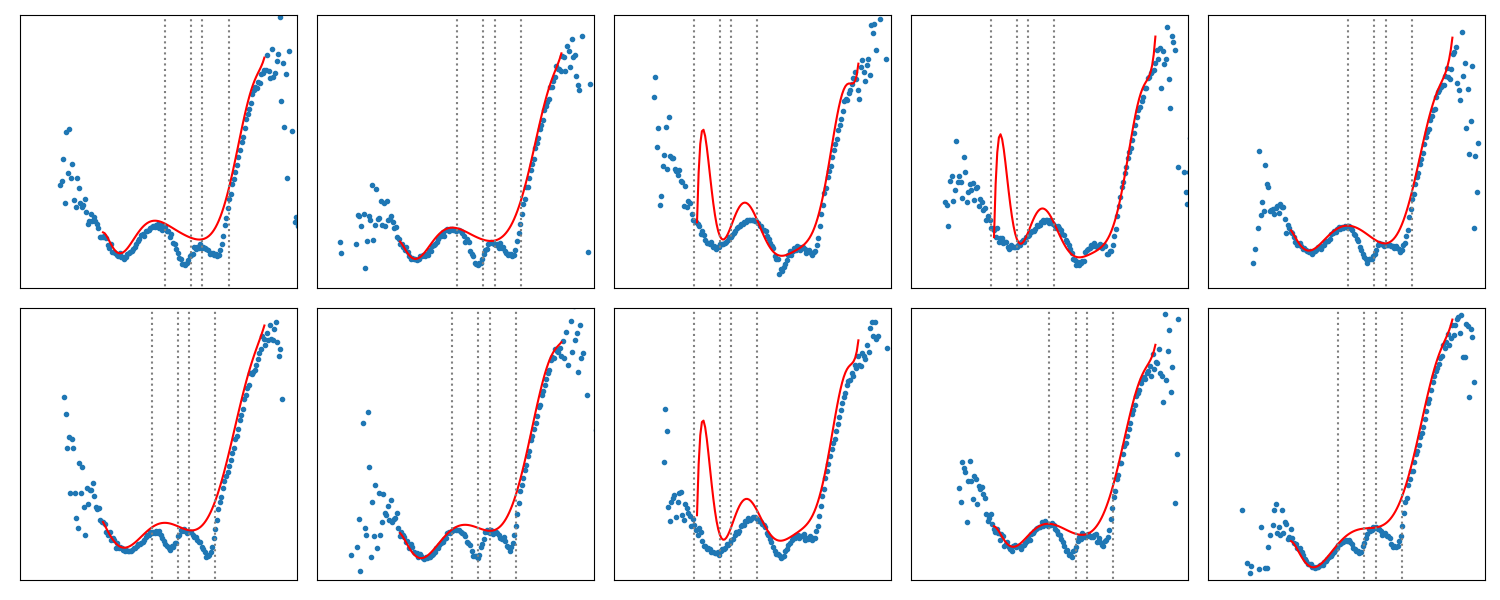}
    \caption{Resulting regression profiles for ten different samples. In case c) and h) the regression predicts additional maxima. In all cases it does not follow the drop completely. Therefore this approach produces a higher spread and an offset compared to the GEM benchmark.}
    \label{fig:polysomeexamples}
\end{figure*}

For the polynomial regression we used an algorithm described in \cite{localreg}. Zeng et al. \cite{zeng2022online} uses a 13 dimensional polynomial, we find that our algorithm does not converge in this case. In Fig. \ref{fig:polyDegree} we show one example of the regression method with increasing polynomial degrees. In our benchmark we use the highest degree where the algorithm converges, with a maximum of 13.
In Fig. \ref{fig:polysomeexamples} we show resulting regressions for ten different samples.
We found that the predictions have a better correlation with the GEM if we ignore the (obviously wrong) parts of the predicted profile that are negative. Still, this method struggles to predict the detailed structure within the masked region, greatly overestimating the influence of the lasing process.

\section{A8: Residual error benchmark}
An additional method to benchmark our method is to evaluate the generated power profiles for data where no lasing process took place, as we have a 'ground truth', meaning the actual value available. Ideally, for this setup, the difference between the XTCAV data and the reconstruction would be zero, resulting in a mean squared error of zero, and also the integral of the power profiles would be zero. For this benchmark we used 460 lasing-off samples  that were not part of the network training (not in the training or validation set) and used each of them 10 times with randomly set masks.

\begin{figure*}[h!b]
    \centering
    \includegraphics[width=0.99\textwidth]{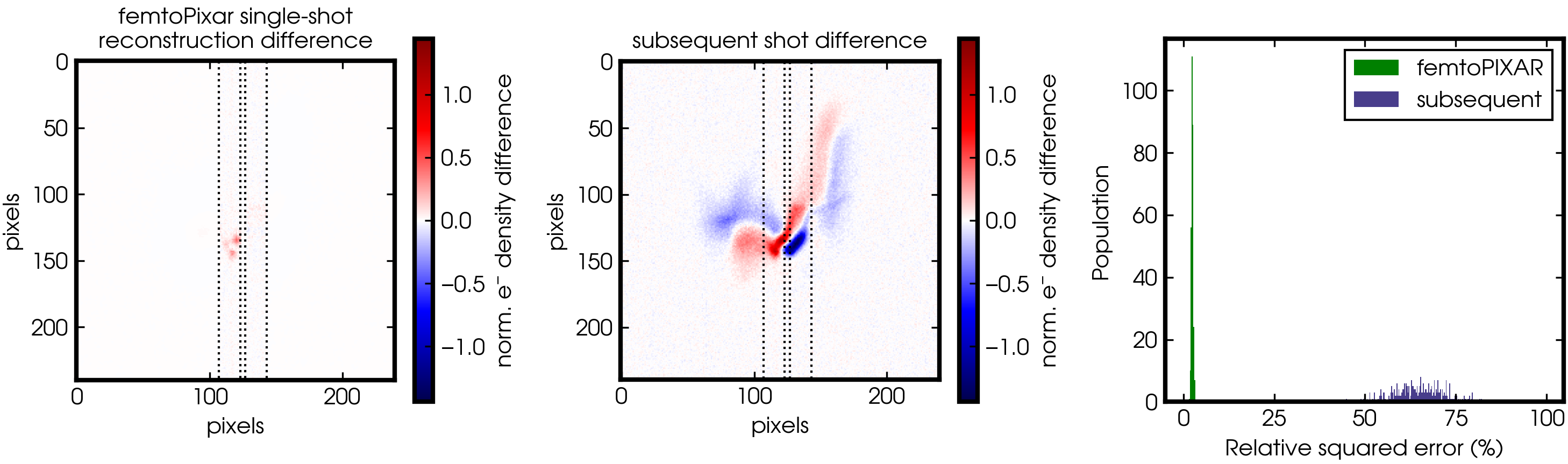}
    \caption{The left image shows an example of the mean error between a lasing-off shot and its reconstruction using the femto-Pixar approach. The middle image shows, for comparison, the difference between a lasing off shot and its subsequent shot. The histogram displays the mean squared error for all data points from the two approaches.}
    \label{fig:diffpixelvalues}
\end{figure*}

Fig. \ref{fig:diffpixelvalues} shows the histogram of relative squared errors for the lasing off reconstructions, evaluated within the masked regions. 
To demonstrate that these values indeed indicate strong similarity, we also show the  histogram of the relative squared difference between the lasing off shots and their respective subsequent shots - where the difference is often not clearly visible by the eye. We show that our approach clearly succeeds in this benchmark. The relative squared error is calculated using the following formula:

\begin{equation}
\text{relative squared error} = \frac{\sum_{x,y} (r_{x,y} - i_{x,y})^2}{\sum_{x,y} i},
\end{equation}
where $r$ and $i$ are once again the reconstructed and input images.

\begin{figure*}[h!b]
    \centering
    \includegraphics[width=0.69\textwidth]{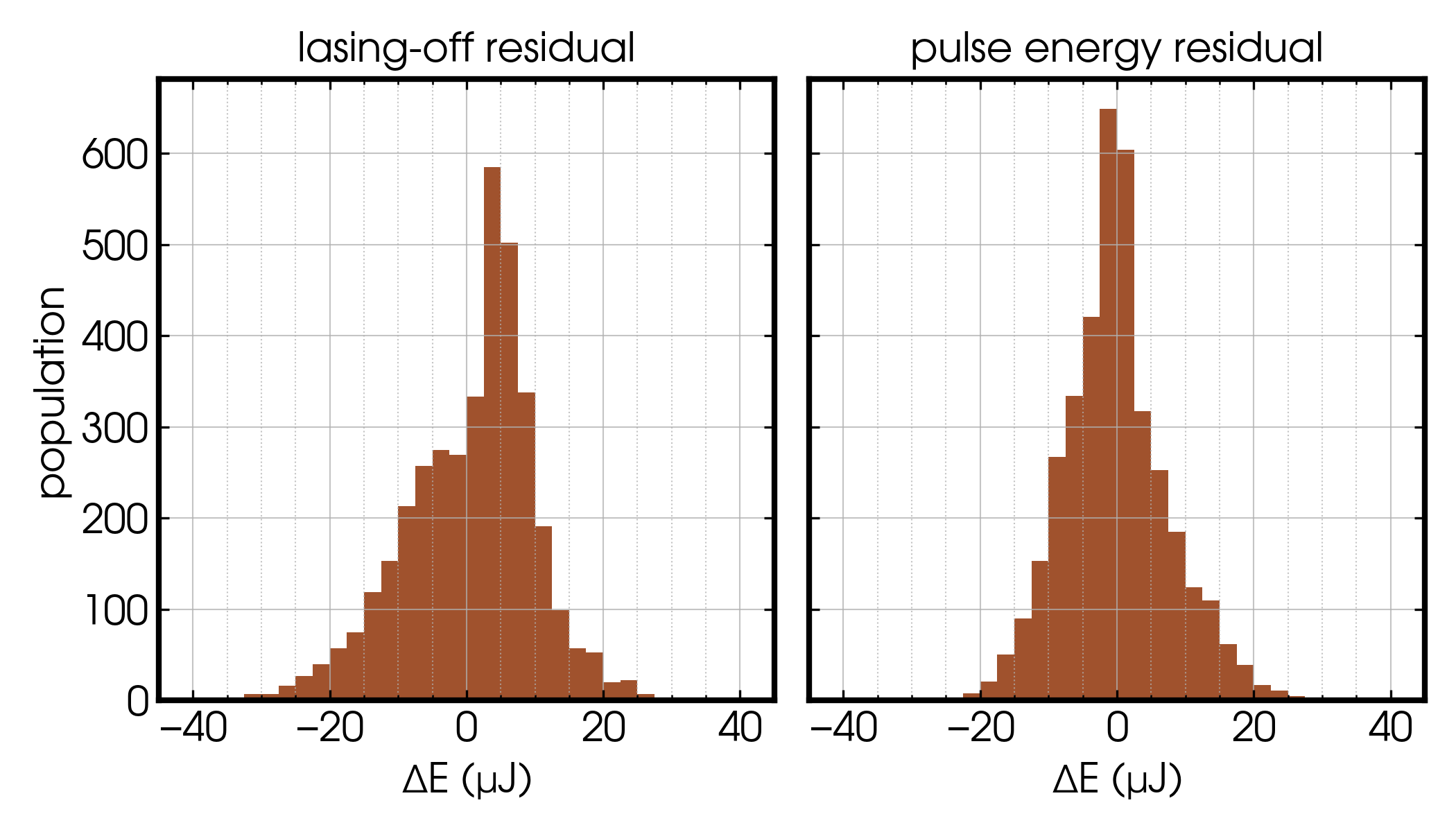}
    \caption{Histogram of the deviation from zero of the derived sum energy of lasing-off power profile predictions. On the left you can see the sum of the integral of the Gaussian fits of both power profiles, on the right the difference between the first and the second peak. The sum is biased towards the positive side as the starting values for the Gaussian fits are positive.}
    \label{fig:diffLoffpeaks}
\end{figure*}

Fig. \ref{fig:diffLoffpeaks} shows a histogram of the resulting pulse energies of 4600 non-lasing data points.  80$\%$ of the data has a deviation from zero of less than 15 $\mu$J, 99.8 $\%$ less than 25 $\mu$J. As we are interested in the pulse ratio between the two peaks it is important that one of the pulses is not systematically under/over estimated with respect the other. We check this on the right side of Fig. \ref{fig:diffLoffpeaks} and see that the pulse energy difference between the two pulses is centered well around zero. This indicates that the pulse ratio is preserved by the reconstruction in this dataset.

\FloatBarrier

\section{A9: Time distribution extracted from reconstructed profiles}
\begin{figure*}[h!bt]
\centering
\includegraphics[width=0.69\textwidth]{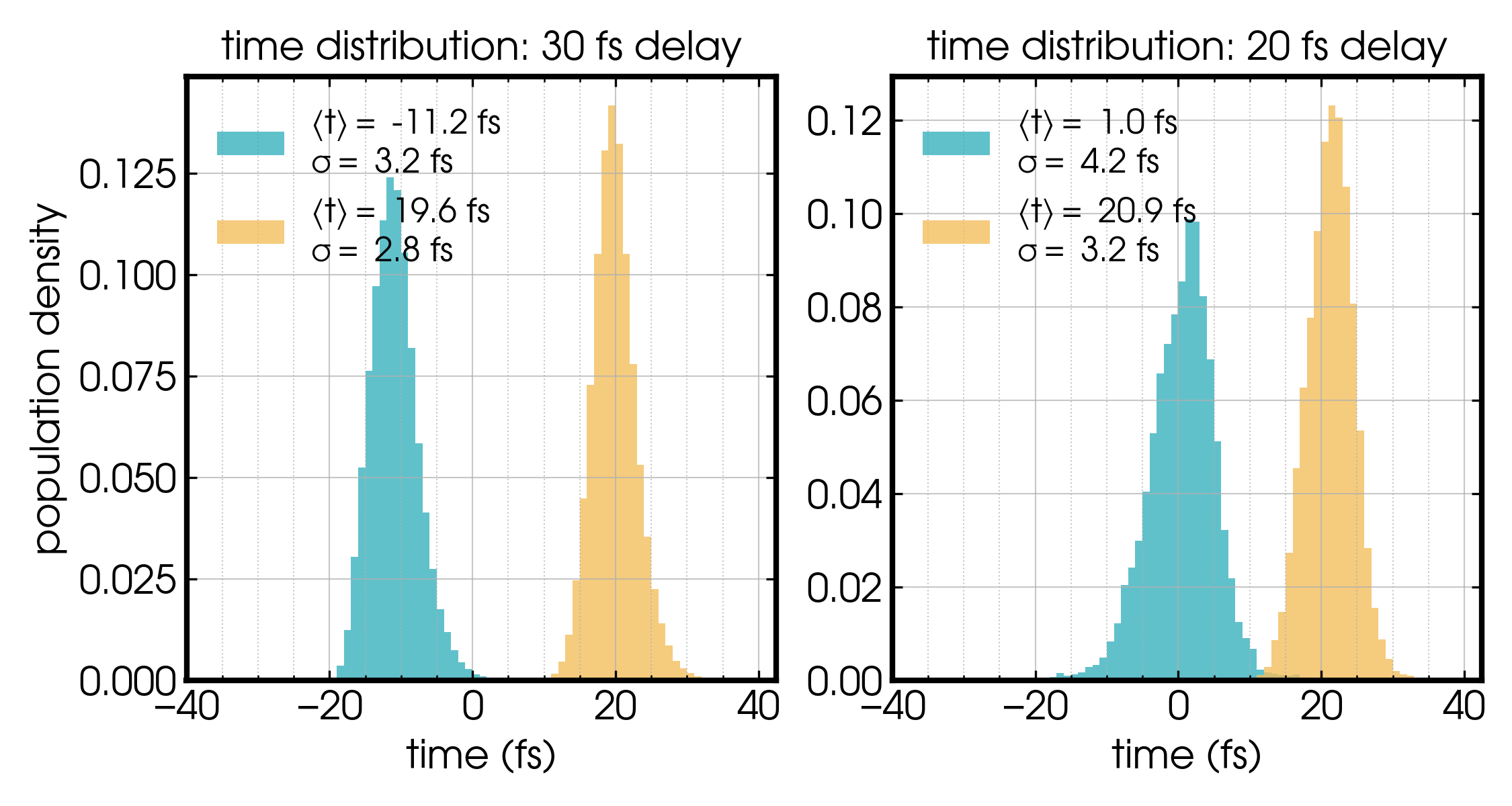}
\caption{Distribution of pulse arrival times from both two-pulse setups. $t=0$ is defined by the center of mass of the XTCAV spectrograph.}
\label{fig:delT}
\end{figure*}
Using a double-Gaussian fit over the reconstructed transient X-ray power profiles $P(t)$, we are able to determine the arrival times of the two pulses at the XTCAV target.  The time axis in \ref{fig:delT} is centered with respect to the COM of the spectrograph in the temporal domain (horizontal axis).  The reconstructed pulses obtained by the Gaussian fits produce bimodal time distributions that match the X-FEL setups, with average time jitter of $\sim 3$~fs that matches well with the 9$\%$ jitter observed by Ding et al. \cite{Ding2015}.

\FloatBarrier
\end{document}